\documentclass[superscriptaddress,twocolumn,preprintnumbers,arxiv,longbibliography]{revtex4-1}
\usepackage{times,xspace}
\usepackage{amsbsy,amssymb,amsmath,bm}
\usepackage{graphicx,color,epsfig}
\usepackage{fancyhdr}
\usepackage{soul}
\usepackage[colorlinks=true,
linkcolor=blue,
urlcolor=blue,
citecolor=blue]{hyperref}

\def\bbbc{{\mathchoice {\setbox0=\hbox{$\displaystyle\rm C$}\hbox{\hbox
to0pt{\kern0.4\wd0\vrule height0.9\ht0\hss}\box0}}
{\setbox0=\hbox{$\textstyle\rm C$}\hbox{\hbox
to0pt{\kern0.4\wd0\vrule height0.9\ht0\hss}\box0}}
{\setbox0=\hbox{$\scriptstyle\rm C$}\hbox{\hbox
to0pt{\kern0.4\wd0\vrule height0.9\ht0\hss}\box0}}
{\setbox0=\hbox{$\scriptscriptstyle\rm C$}\hbox{\hbox
to0pt{\kern0.4\wd0\vrule height0.9\ht0\hss}\box0}}}}

\newcommand{\ignore}[1]{}
\newcommand{\mComment}[1]{}
\newcommand{\gComment}[1]{}
\newcommand{\jComment}[1]{}
\newcommand{\rComment}[1]{}
\newcommand{\lComment}[1]{}

\renewcommand{\gComment}[1]{\textcolor{magenta}{Gerardo: #1}}
\begin{document}
\title{Exciton condensation in bilayer spin-orbit insulator}
\author{Hidemaro Suwa}
\affiliation{Department of Physics, The University of Tokyo, Tokyo 113-0033, Japan}
\affiliation{Department of Physics and Astronomy, The University of Tennessee,
Knoxville, Tennessee 37996, USA}
\author{Shang-Shun Zhang}
\affiliation{Department of Physics and Astronomy, The University of Tennessee,
Knoxville, Tennessee 37996, USA}
\author{Cristian D. Batista}
\affiliation{Department of Physics and Astronomy, The University of Tennessee,
Knoxville, Tennessee 37996, USA}
\affiliation{Quantum Condensed Matter Division and Shull-Wollan Center, Oak Ridge National Laboratory, Oak Ridge, Tennessee 37831, USA}

\date{\today}
\begin{abstract}
We investigate the nature of the magnetic excitations of a bilayer single-orbital Hubbard model in the intermediate-coupling regime.  This model exhibits a quantum phase transition (QPT) between a paramagnetic (PM) and an insulating antiferromagnetic (AFM) phase at a critical value of the coupling strength.
By using the random phase approximation, we show that the QPT is continuous when the PM state is a band insulator and that the corresponding quantum critical point (QCP) arises from the condensation of preformed excitons. 
These low-energy excitons reemerge on the other side of the QCP as the transverse and longitudinal modes of the AFM state.
In particular, the longitudinal mode remains sharp   for the model parameters relevant to Sr$_{3}$Ir$_{2}$O$_{7}$ because of the strong easy-axis anisotropy of this material.  
\end{abstract}

\maketitle

\section{Introduction}
Low-dimensional Mott insulators play a central role in correlated electron physics because of the
novel  states of matter and phase transitions that they can host.
In the strong-coupling limit, $U/\vert t \vert \gg 1$, the on-site repulsion $U$ ``freezes out'' the charge degrees of freedom, turning the Mott insulator  into a quantum magnet described  by an effective spin Hamiltonian. 
The lattice connectivity  can then  be exploited to generate competing exchange interactions that induce quantum phase transitions (QPTs) between different states of matter. 
One of the simplest examples is provided by bilayer  materials~\cite{Jaime04,Sebastian06,Batista07,Zapf14,Stone08,Kofu09},
where the competition between intralayer and interlayer hopping amplitudes, $t$ and $t_z$, can induce a transition between an antiferromagnetic (AFM) state and  a quantum paramagnet (QPM) comprising  local singlet states on the interlayer  dimers. This QPT, which is typically induced by applying pressure, has been extensively studied in multiple quantum magnets  to understand different properties of the  QCP, such as the emergence and  decay of the longitudinal mode (LM) that is present in the AFM state~\cite{chubukov1994,sachdev1999,Matsumoto04,Ruegg08,Podolsky11,Podolsky12,Lohoefer15,Hong16,Qin17,Su20}.

The discovery of  low-dimensional intermediate-coupling  4$d$ and 5$d$-electron correlated insulators introduces a knob, $U/t$, that can  be used to unleash the charge degrees of freedom via pressure or strain. 
For instance, the iridate materials~\cite{Arita13,Rau16,Cao18,Hao18,Hao19,Bertinshaw19} have a charge gap $\Delta$  that is comparable to the magnon bandwidth $W$. The reduction of $U/t$ leads to the suppression of the AFM ordering in favor 
of a  paramagnetic state.
In bilayer materials, such as Sr$_3$Ir$_2$O$_7$, the paramagnetic state can either be metallic or a band insulator depending on the spin orbit coupling.
Similarly to the large $U/\vert t \vert$ limit, where the transition from the QPM to the  AFM state can be described as a triplon condensation, 
 the QPT from the band insulator to the AFM state corresponds to condensation of preformed magnetic excitons. These bound states reemerge on the other side of the QCP as longitudinal and transverse modes (magnons) of the AFM state. In contrast, the longitudinal mode (LM) is absent in the AFM phase induced by a metal-insulator transition  with perfect Fermi surface (FS) nesting.

In this article, we study the possible  QPTs induced by reducing the  $U/\vert t \vert$ ratio in bilayer materials.
Although the transition into the band insulator has some similarities with the transition into the QPM of pure spin systems ($U/\vert t \vert \gg 1$)~\cite{chubukov1994,sachdev1999,Matsumoto04,Ruegg08,Podolsky11,Podolsky12,Lohoefer15,Hong16,Qin17,Su20},  there are also some important differences associated with the enhanced charge fluctuations.
By applying our analysis to the easy-axis bilayer antiferromagnet Sr$_{3}$Ir$_{2}$O$_{7}$, with $\Delta=130$meV~\cite{Okada13} and $W=70$meV~\cite{Kim12}, we reveal the existence of a LM in  {\it some regions of the Brillouin zone}, which arises from the band-insulating character of the noninteracting limit of the model: The N{\'e}el phase  is induced by condensation of  preformed $S^z=0$ excitons at a critical coupling strength  $U=U_{\rm c}$. This exciton  reemerges in the N{\'e}el phase ($U>U_{\rm c}$) as a LM
whose energy scale and dispersion are consistent with previous resonant inelastic x-ray scattering (RIXS) measurements~\cite{Kim12,Moretti15,Hogan16,Lu17}.  It is important to note that the LM that emerges near the same QCP  in the strong coupling limit of the Hubbard model~\cite{Moretti15} exists {\it over the whole Brillouin zone} and its energy is much lower than the charge gap because it is obtained from a pure spin model. In contrast, the LM that we are proposing for Sr$_{3}$Ir$_{2}$O$_{7}$  only exists in finite regions of momentum space, around the wave vectors ${\bm q}={\bm 0}$ and ${\bm q}={(\pi,\pi)}$, because it is induced by strong charge fluctuations  ($U \sim U_{\rm c}$). Indeed, the mode disappears inside the particle-hole continuum for wave vectors that are far enough from  ${\bm q}={\bm 0}$ and ${\bm q}={(\pi,\pi)}$.

Our results then suggest that Sr$_{3}$Ir$_{2}$O$_{7}$ is a realization of the long sought excitonic insulator that was predicted almost 60 years ago~\cite{Cloizeaux65,Halperin68,Jerome67}.
However, higher resolution RIXS experiments are needed to confirm this prediction.

\section{Model}
We consider a bilayer single-orbital Hubbard model $\mathcal{H} =- \mathcal{H}_{\rm K} +\mathcal{H}_{{\rm U}}$, with
$\mathcal{H}_{\rm U} \! =  U\sum_{\bm r} n_{\bm{r} {\uparrow}}  n_{\bm{r} {\downarrow}}$ and
\begin{eqnarray}
\!\!\! \mathcal{H}_{\rm K}  \! =  \!
t \!  \sum_{{\bm r},\nu} \! {\bm c}^{\dagger}_{\bm r}
{\bm c}_{{\bm r} + {\bm a}_{\nu}} \!\! + t_z \!\! \sum_{{\bm r}_{\bot}} \!
 {\bm c}^{\dagger}_{({\bm r}_{\bot},1)} e^{i\frac{\alpha}{2}\epsilon_{\bm r}\sigma_{z}}
{\bm c}_{({\bm r}_{\bot},2)}
\! \!+ \! {\rm H. c.},
\label{eq:H_bl}
\end{eqnarray}
where $t,t_z,\alpha,U \in \mathbb{R}$, ${\bm c}^{\dagger}_{\bm r} \equiv [ c^{\dagger}_{\uparrow, {\bm r}}, c^{\dagger}_{\downarrow, {\bm r}}]$, is the Nambu spinor of the electron field [${\bm r}\equiv({\bm r}_{\bot},l)$, $l=1,2$ is the layer index and ${\bm r}_{\bot}=r_1{\bm a}_1 + r_2 {\bm a}_2$], and ${\bm a}_{\nu}$ ($\nu=1,2$) are the primitive vectors of the square lattice of each layer. The sign $\epsilon_{\bm r}$ takes the values 1 ($-1$) for ${\bm r} \in \mathcal{A}$ ($\mathcal{ B}$) sublattice of the bipartite bilayer system.
This Hamiltonian is an effective model for a bilayer system with finite  SOC, which is realized in ruthenates and iridates. For example, the large SOC of the bilayer iridates splits the 5$d$ $t_{2g}$ orbitals of the Ir$^{4+}$ ion  into $J=1/2$ and $3/2$ multiplets~\cite{Kim08}.  Consequently, $\mathcal{H}$ becomes a low-energy model for the $5d$ hole ($5d^5$ electronic configuration) of the strontium iridates after projecting the relevant multi orbital Hubbard model onto the lowest energy $J=1/2$ doublet. The phase $\alpha$ arises from hopping matrix elements between $d_{xz}$ and $d_{yz}$ orbitals allowed by staggered octahedral rotations. The phase of the intralayer hopping is gauged away by applying a sublattice-dependent gauge transformation.

\section{Mean field approximation}
For $\alpha \neq 0$, the model~(\ref{eq:H_bl}) has an easy $z$-axis spin anisotropy  and the ground state of
the AFM phase can have N\'{e}el ordering, $\langle S^{\mu}_{\bm{r}}\rangle = (-1)^{\gamma_{\bm r}}M\delta_{\mu z}$, where $\gamma_{\bm r}=(1 + \epsilon_{\bm r})/2$,
${S}^{\mu}_{\bm{r}} = {1/2}c^{\dagger}_{\bm{r}} \sigma^{\mu}c_{\bm{r}}$ $(\mu=x,y,z)$, and $M$ is the magnetization. 
A mean field decoupling of $\mathcal{H}_{\rm U}$ leads to
\begin{equation}
\! {\cal H}_{\rm U}^{\rm MF} \! = \! - U M\sum_{\bm{r}}(-1)^{\gamma_{\bm r}}c^{\dagger}_{\bm{r}}\sigma_{z}c_{\bm{r}} + {1\over 2}Un\sum_{{\bm r}} c^{\dagger}_{\bm{r}} c_{\bm{r}}  + C,
\end{equation}
where $C=U M^{2}{\cal N}_{s}  - {1\over 4}Un^2 {\cal N}_{s}$, ${\cal N}_{s}$ is the number of lattice sites, $n={\cal N}_s^{-1}\sum_{\bm r}\langle c^{\dagger}_{\bm{r}} c_{\bm{r}} \rangle$ is the electron density 
and the second term can be absorbed into the chemical potential.

\begin{figure}
\includegraphics[width=\columnwidth]{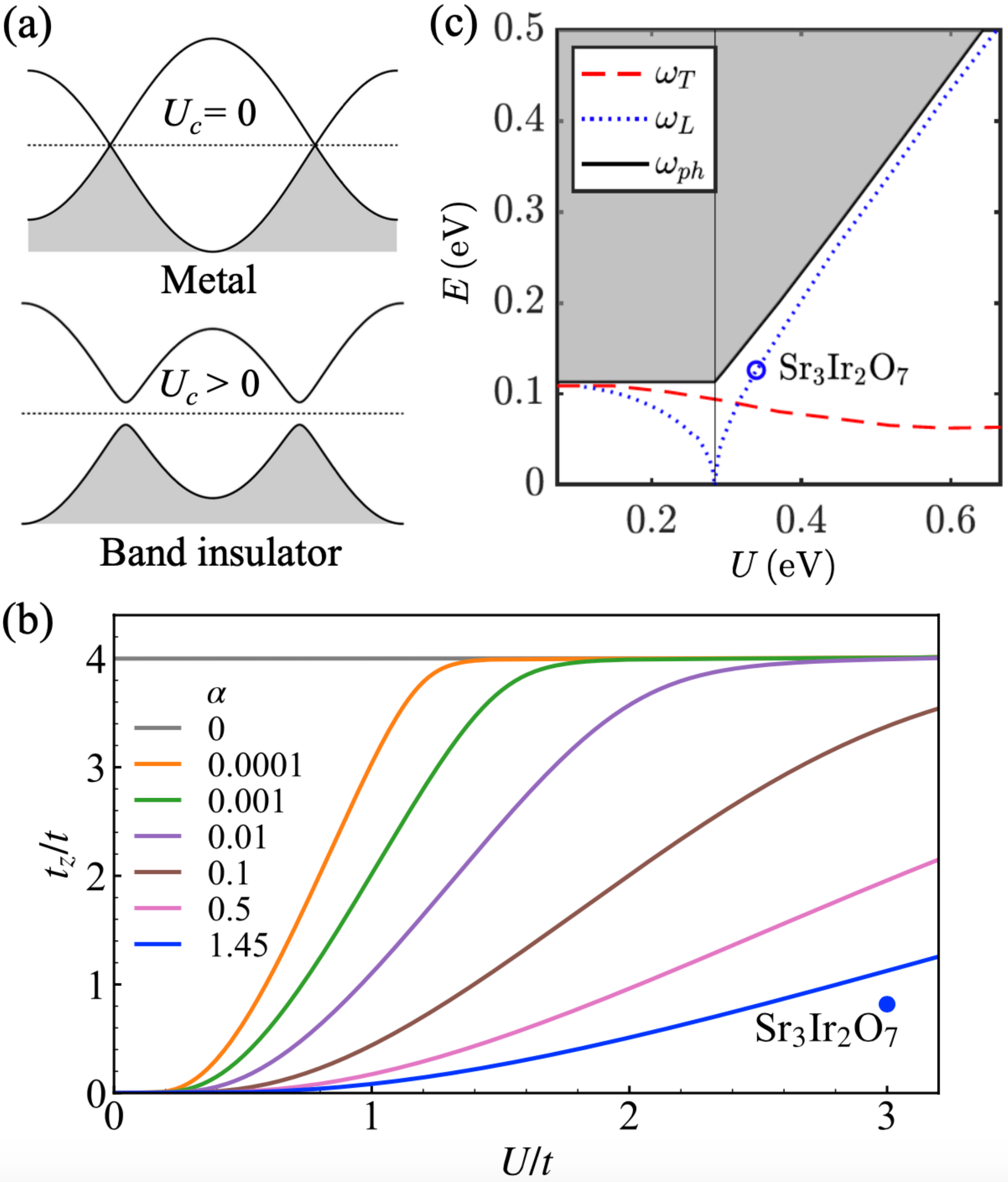}    
\caption{ (a) Schematic picture for the band structure of noninteracting metal and band insulator.
  (b) Phase boundaries in the mean field approximation for $\alpha=0, 10^{-4}, 10^{-3}, 10^{-2}, 10^{-1}$, 0.5, and 1.45 (from top to bottom). For $\alpha = 0$ and $U=0$, the system is a metal for $|t_z/t|<4$ and a band insulator for $|t_z/t|>4$. The metal becomes an AFM insulator for an infinitesimally small $U$. For $\alpha \neq 0$, the present model~\eqref{eq:H_bl} exhibits a continuous phase transition from the band insulator to the AFM insulator at a finite $U/t$. The solid circle symbol indicates the estimated parameter set for the bilayer iridate Sr$_3$Ir$_2$O$_7$ with $\alpha=1.45$.
  (c) Energy diagram as a function of $U$ for the parameter set relevant to Sr$_3$Ir$_2$O$_7$. The red dashed line indicates the energy of the $S^z=\pm1$ exciton that becomes the transverse mode for $U>U_{\rm c}$. The  blue dotted line indicates the energy of the $S^z=0$ exciton that becomes a longitudinal mode for $U>U_{\rm c}$. The gray area indicates the particle-hole continuum.
From a comparison with RIXS data~\cite{Kim12,Moretti15,Hogan16,Lu17}, we estimate that $U=0.33$\,eV for Sr$_3$Ir$_2$O$_7$.
}
\label{fig:gap}
\end{figure}

By Fourier transforming annihilation and creation operators,
\begin{equation}
\! c_{\sigma,\bm{r}} \! = \! \frac{1}{\sqrt{{\cal N}_{u}}}\sum_{\bm{k} \in {\rm BZ}}e^{i(\bm{k}_{\bot}\cdot\bm{r}_{\bot}+k_{z}l)}c_{\gamma\sigma,\bm{k}}, \ {\bm r}\in \text{sublattice }\gamma,
\end{equation}
where ${\cal N}_{u}=\frac{1}{2}{\cal N}_{s}$ is the number of unit
cells, and  ${\bm k} \equiv ({\bm k}_{\bot},k_z)$ runs over the first Brillouin zone (BZ): ${\bm k}_{\bot}=k_1{\bm b}_1^{\prime}+k_2{\bm b}_2^{\prime}$, with ${\bm b}_1^{\prime}=(1/2,-1/2)$, ${\bm b}_2^{\prime}=(1/2,1/2)$, {$k_1,k_2 \in [0,2\pi)$} and $k_z=0,\pi$,  we obtain the momentum space representation of
\begin{equation}
{\cal H}^{\rm MF}
= \sum_{\bm{k}\in {\rm BZ}}  {\bm c}^{\dagger}_{\bm k} {\cal H}^{\rm MF}_{\bm{k}} {\bm c}^{\;}_{\bm k}, 
\end{equation}
with
\begin{align}
{\cal H}^{\rm MF}_{\bm{k}} & =
\left(\begin{array}{cc}
UM\sigma_{z} & \epsilon_{\bm{k}}^{(1)}  -t_{z}\cos(k_{z})e^{-i\frac{\alpha}{2}\sigma_{z}}\\
\epsilon_{\bm{k}}^{(1)}  -t_{z}\cos(k_{z})e^{i\frac{\alpha}{2}\sigma_{z}} & - UM \sigma_{z}
\end{array}\right), \nonumber
\end{align}
${\bm c}_{\bm{k}} \equiv (c_{\mathcal{A} \uparrow,\bm{k}},c_{\mathcal{A} \downarrow,\bm{k}},c_{\mathcal{B} \uparrow,\bm{k}},c_{\mathcal{B} \downarrow,\bm{k}})^T$ and 
\begin{equation}
\epsilon_{\bm{k}}^{(1)} = -2t \left( \cos {k_1+k_2 \over 2} + \cos{k_1-k_2 \over 2} \right).
\end{equation}
${\cal H}^{\rm MF}_{\bm{k}}$ is diagonalized by a unitary $4 \times 4$  matrix  $U(\bm{k})$,
\begin{eqnarray}
c_{\gamma\sigma,\bm{k}} & = & \sum_{n}U_{(\gamma\sigma),n}(\bm{k})\psi_{n,\bm{k}},
\end{eqnarray}
where $n\equiv(s,\sigma)$ with $s=\pm$, $\sigma=\uparrow,\downarrow$ and
each column of  $U(\bm{k})$ is an eigenvector of ${\cal H}^{\rm MF}_{\bm{k}}$:
\begin{eqnarray}
X_{s \uparrow}(\bm{k}) \! = \!
\left(\begin{array}{c}
x_{\uparrow\bm{k} }\sqrt{\frac{1 + s z_{\uparrow\bm{k} }}{2}}\\
0\\
s \sqrt{\frac{1 - s z_{\uparrow \bm{k} }}{2}}\\
0
\end{array}\right), 
X_{s\downarrow}(\bm{k} ) \! = \! \left(\begin{array}{c}
0\\
x_{\downarrow\bm{k} }\sqrt{\frac{1-s z_{\downarrow\bm{k} }}{2}}\\
0\\
s\sqrt{\frac{1 + s z_{\downarrow\bm{k} }}{2}}
\end{array}\right), \quad \ 
\end{eqnarray}
with
\begin{eqnarray}
 x_{\sigma\bm{k} } &=& \frac{b_{\sigma\bm{k} }}{\rvert b_{\sigma\bm{k} }\rvert}, \;\;z_{\sigma\bm{k} }=\frac{\delta}{\sqrt{\delta^{2}+\rvert b_{\sigma\bm{k} }\rvert^{2}}}, 
\nonumber \\
b_{\sigma\bm{k} } &=& \epsilon_{\bm{k}}^{(1)} -t_{z}\cos(k_{z})e^{-i\sigma\frac{\alpha}{2}},
\end{eqnarray}
and $\delta=U M$. The corresponding eigenenergies,
$
\varepsilon_{s \sigma}(\bm{k} ) = s  \sqrt{\delta^{2}+\rvert b_{\sigma\bm{k} }\rvert^{2}}, 
$
are  independent of the spin flavor, $\rvert b_{\sigma\bm{k} }\rvert^{2} = b_{\bm k}^2$ and $z_{\sigma\bm{k} } \equiv z_{\bm k}$, because of the U(1) invariance of ${\cal H}$ under global spin rotations about the $z$ axis. The noninteracting system is then metallic when  the band gap at each ${\bm k}$ point, $\Delta_{\bm k}=2\sqrt{\delta^2+b_{\bm k}^2}=2|b_{\bm k}|$, closes on a nodal line $b_{\bm k}=0$ of band crossing points that coincides with the FS [see Fig.~\ref{fig:gap}(a)]. 

Because we are considering a bilayer system, the half-filled condition corresponds to an {\it integer number} of electrons per unit cell, implying that the ground state of the noninteracting ($U=0$) system can either be a {\it metal} or a {\it band insulator}, as illustrated in  Fig.~\ref{fig:gap}(a). For $\alpha=0$, the metal to insulator transition occurs via a semimetallic state with small and identical electron and hole pockets that shrink into a quadratic Fermi point.  
For instance, for $t_z=0$, the FS is the square defined by the equations $k_1+k_2=\pm\pi$ and $k_1-k_2=\pm \pi$.
Upon increasing $\rvert t_z \rvert$, while keeping $\alpha=0$, the square shrinks  into a circular pocket that finally  disappears for $\rvert t_z \rvert > 4 \rvert t \rvert$.
In contrast, the noninteracting system is always a band insulator for $\alpha \neq 0$.

The on-site repulsion $U$  induces AFM ordering via a QPT that depends on the nature of the noninteracting state.
In both cases, noninteracting metal and insulator, the longitudinal AFM susceptibility (${\bm q}={\bm 0}$) at $\omega=0$  is given by
\begin{eqnarray}
 \chi_0^{(\gamma_1,z);(\gamma_2,z)} = (-1)^{\gamma_1+\gamma_2} {1\over 2{\cal N}_u } \sum_{{\bm k}\in {\rm BZ}} {1\over \rvert b_{\bm k} \rvert},
 \end{eqnarray} 
where $\gamma_{1,2}={\cal A},{\cal B}$ are sublattice indices.
Note that the  antiferromagnetic susceptibility can diverge at ${{\bm q}={\bm 0}}$ because the nonmagnetic unit cell contains one site of each sublattice. From the mean field calculation, the critical interaction strength is given by
\begin{eqnarray}\label{eq:uc}
U_{\rm c}^{-1} & = & {1\over 2{\cal N}_u } \sum_{{\bm k}\in {\rm BZ}} \frac{1}{\rvert b_{\bm k} \rvert}.
\end{eqnarray}
The resultant phase boundaries for several values of $\alpha$ are shown  in Fig.~\ref{fig:gap}(b). 

For $\alpha=0$ and $U=0$, the system is metallic for $|t_z/t|<4$. In this case, the perfect nesting due to the coincidence of the particle and hole Fermi surfaces leads to a logarithmic divergence of $ \chi_0^{(\gamma_1,z);(\gamma_2,z)}({\bm q}, \omega=0)$ at ${{\bm q}={\bm 0}}$. Note that the divergence becomes $\ln^2{q}$ for a square FS ($t_z=0$) because of the Van Hove singularity in the density of states at the Fermi level (the corners of the square correspond to saddle points of the dispersion relation).
In other words, the metal becomes an AFM insulator for an infinitesimally small value of $U$, implying that the critical interaction strength $U_{\rm c}$ is equal to zero.
In contrast,  the system becomes a band insulator (finite charge gap) for $\alpha \neq 0$ because the sublattice symmetry is no longer present. As shown in Fig.~\ref{fig:gap}(b), the metal-insulator transition at $U=0$ is then replaced by a continuous QPT between the band and the AFM insulators at a finite $U$ value,  $U=U_{\rm c}$,  because the integral \eqref{eq:uc} is now convergent. This phase diagram clearly shows the significance of the spin-orbit coupling: The proximity to the critical point is caused by a finite $\alpha$ even for small $t_z/t$. The physical interpretation of the difference between the metallic and the band-insulating cases will become clearer upon analyzing the  behavior of the magnetic excitations.

The order parameter is determined by solving the self-consistent mean field equation 
\begin{equation}
M=\frac{1}{2{\cal N}_{u}}\sum_{\bm{k}} z_{\bm{k}}.
\end{equation}
Near the critical point $U=U_{\rm c}$, we obtain 
$M\propto (t/U)e^{-1/(\rho_0 U)}$ for the metal, where $\rho_0$ the density of states at the Fermi level,
and $M=2\sqrt{(U-U_{\rm c})/(DU_{\rm c}^4)}$ for the band insulator with
$D= {1\over {\cal N}_u } \sum_{{\bm k}\in {\rm BZ}} {\rvert b_{\bm k} \rvert^{-3}}$.

\section{Exciton condensation}
In view of the $U(1)$ invariance of ${\cal H}$  and  the AFM ground state, the transverse and longitudinal spin fluctuations are decoupled from each other. Within the random phase approximation (RPA), the magnetic susceptibilities of the transverse and the longitudinal modes are given by
\begin{eqnarray}
\chi^{+-}(\bm{q},i\omega_{n}) & = & \frac{1}{ \tau^{0}-U\chi_{0}^{+-}(\bm{q},i\omega_{n})}\chi_{0}^{+-}(\bm{q},i\omega_{n}),\label{eq:Trans_pm}\\
\chi^{zz}(\bm{q},i\omega_{n}) & = & \frac{1}{\tau^{0}-{U\over 2}\chi_{0}^{zz}(\bm{q},i\omega_{n})}\chi_{0}^{zz}(\bm{q},i\omega_{n}) \label{eq:rpa_L},
\end{eqnarray}
respectively, where $\tau_0$ is the $2\times 2$ identity matrix. Here $\chi^{+-}$ and $\chi^{zz}$ refer to $2\times2$ matrices in the sublattice space, while $\chi_{0}^{+-}$ and $\chi_{0}^{zz}$ are the  bare magnetic susceptibilities. See Appendix~\ref{rpa} for details of the RPA calculation.

The eigenfrequencies $\omega_{\bm q}$  of the collective transverse modes (magnons) can be extracted from the poles of the transverse RPA susceptibility: $\det[\tau^{0}-U\chi_{0}^{+-}(\bm{q},\omega_{\bm q})] = 0$. 
The spectrum  is fully gapped because the U(1) symmetry of ${\cal H}$ 
is not spontaneously broken by the  AFM ordering along the $z$-axis. 
As shown in Fig.~\ref{fig:gap}(c), the transverse modes remain gapped at $U=U_{\rm c}$ and they become the  $S^z=\pm1$ exciton modes of nonmagnetic band insulator for $U<U_{\rm c}$.
Note that this gap closes in the absence of  spin-orbit coupling ($\alpha=0$) because the Hamiltonian becomes SU(2) invariant.

The most interesting feature is the emergence of a LM  below the particle-hole 
continuum around the $\Gamma$ point of the second Brillouin zone [${\bm q}=(\pi,\pi,\pi)$]. The origin of this mode can be understood by analyzing the excitation spectrum of the  band insulator for $U<U_{\rm c}$.
The bare magnetic susceptibility at ${\bm q}={\boldsymbol 0}$, which is equivalent to $(\pi,\pi,\pi)$,  is 
$\chi_{0}^{zz}(\bm{0},\omega) = \Phi( \omega) (\tau^0 - \tau^x)$, where $\tau^x$ is the Pauli matrix and
\begin{equation}
\Phi(\omega)  =\frac{1}{2{\cal N}_{u}}\sum_{\bm{k} \in {\rm BZ}} {1\over \rvert b_{\bm{k}}\rvert } \frac{1}{1-(\frac{\omega}{{2b_{\bm k}}} )^{2} }.
\end{equation}
The function $\Phi(\omega)$ is real, and it increases monotonically with $\omega$ up to the lower edge of the particle-hole continuum $\omega_{ph}=2\text{min}(b_{{\bm k}})$, where it diverges: $\Phi(\omega=0)=1/U_{\rm c}$ and $\lim_{\omega \to \omega_{ph} } \Phi(\omega)= \infty$. The putative pole of the longitudinal susceptibility  is determined by the condition
\begin{eqnarray}\label{eq:pole_L}
\Phi(\omega)=1/U,
\end{eqnarray}
which implies that a pole must exist in the energy window $0 \leq \omega \leq \omega_{ph}$ for $0<U<U_{\rm c}$.
As long as $U_{\rm c}$ is finite, i.e., the noninteracting system is a band insulator, the pole appears just below the gap
$\Delta=\omega_{ph}$ that signals the onset of the particle-hole continuum. This pole corresponds to the formation of an $S^z=0$ exciton. Upon examining the transverse susceptibility, we also find a doubly degenerate pole associated with the formation of $S^z=\pm 1$ excitons [see Fig.~\ref{fig:gap}(c)], whose energy is higher than the energy of the $S^z=0$ exciton because of the easy-axis anisotropy. We note that the three excitonic states become degenerate in the absence of spin-orbit interaction because ${\cal H}$ is SU(2) invariant in that limit. 
The binding energy exhibits the singular behavior $E_b \propto  \omega_{ph} \exp(-\kappa\omega_{ph}/U)$,  characteristic of 2D systems in the weak-coupling limit ($\kappa$ is a nonuniversal number). 
As shown in Fig.~\ref{fig:gap}(c), the  $S^z=0$ exciton  becomes soft at $U=U_{\rm c}$ and
$\omega_L \simeq 2\sqrt{2(U_{\rm c}-U) /(D U_{\rm c}^2)}$ for $U \simeq U_{\rm c}$. The condensation of this mode signals the onset of 
the N{\'e}el  phase with magnetic moments pointing along the $z$ axis due to the effective easy-axis anisotropy.

\begin{figure*}
\includegraphics[width=2.08\columnwidth]{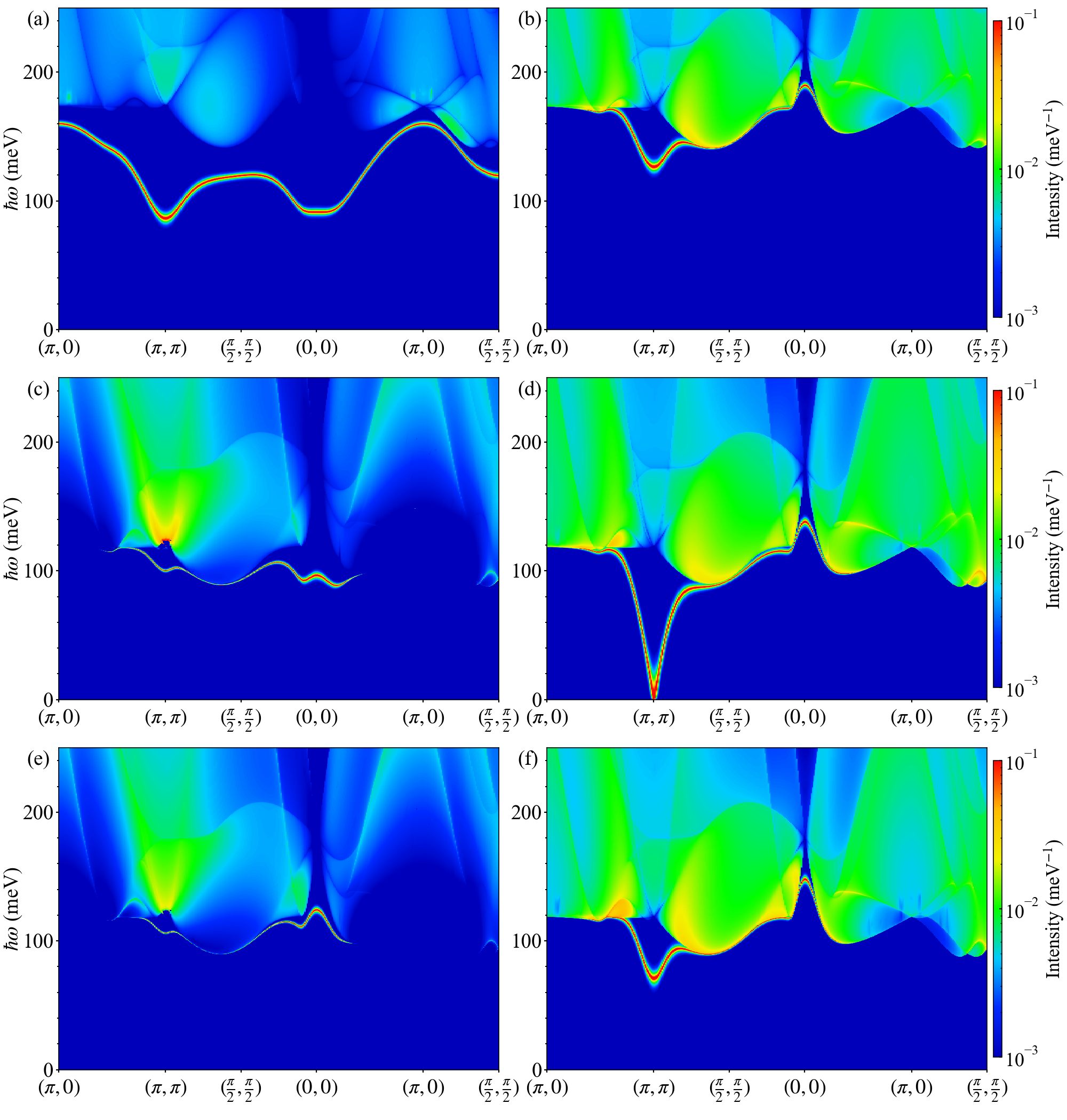}    
\caption{\label{fig:sqo} Dynamical spin structure factors of the out-of-phase mode: (a)~in-plane or transverse
$S^{xx}({\bm q}, \omega) = S^{yy}({\bm q}, \omega)$ component and (b)~out-of-plane or longitudinal component
$S^{zz}({\bm q}, \omega)$ for $U=0.33$~eV; (c) $S^{xx}({\bm q}, \omega) = S^{yy}({\bm q}, \omega)$ and (d) $S^{zz}({\bm q}, \omega)$ for $U=U_{\rm c}\approx0.28$~eV; and (e) $S^{xx}({\bm q}, \omega) = S^{yy}({\bm q}, \omega)$ and (f) $S^{zz}({\bm q}, \omega)$ for $U=0.23$~eV. The hopping parameters were chosen to reproduce the RIXS data of the bilayer iridate Sr$_3$Ir$_2$O$_7$~\cite{Kim12,Moretti15,Hogan16,Lu17}: $t=0.11$~eV, $t_z=0.09$~eV, $\alpha=1.45$, and the intralayer next nearest neighbor hopping $0.012$~eV. The broadening factor is $\eta=10^{-4}$~eV. Intensity larger than the maximum value in the scale (color) bar is plotted in the same color.
}
\end{figure*}
In the AFM phase ($U>U_{\rm c}$), the bare magnetic susceptibility at ${\bm q}={\boldsymbol 0}$  is given by
\begin{align}
\Phi(\omega) & = {1\over 2UM} \frac{1}{{\cal N}_{u}}\sum_{\bm{k} \in {\rm BZ}}z_{\bm{k} }\frac{1-z_{\bm{k} }^{2}}{1-(\frac{\omega}{{2 U M}})^{2}z_{\bm{k} }^{2}}.
\end{align}
Once again,  $\Phi(\omega)$ is real, and $d\Phi/d\omega \geq 0$ for 
 $\omega \leq \omega_{ph}= 2\sqrt{(UM)^2+{\text{min}}(b_{{\bm k}})^2}$. 
If the noninteracting state is a band insulator (finite $U_{\rm c}$), it holds that $b_{\bm k}>0$ for any ${\bm k}$, implying that a pole must then exist in the  window $0 \leq \omega \leq \omega_{ph}$ because $\Phi(\omega=0) < 1/U$ and $\Phi(\omega_{ph}) > 1/U$.  For $U \gtrsim U_{\rm c}$, the exciton energy
scales as $\omega_L=4\sqrt{{(U-U_{\rm c})/(D U_{\rm c}^2)}}$. Upon further increasing $U$, $\omega_L$ increases quickly and approaches the lower edge of the particle-hole continuum asymptotically. 
This behavior can be understood by considering  the large-$U$ limit. In this limit, the effective  particle-hole interaction that provides the ``glue'' for the formation of the LM is the exchange interaction, on the order of $t_z^2/U$, along the vertical bonds, implying that the binding energy of the LM  must vanish for $U \to \infty$.
The particle and the hole break only one AFM link when they occupy the two sites of the same vertical bond, while they break two vertical links when they occupy two different vertical bonds.
In contrast, the binding energy of the low-energy transverse modes is on the order of $U$ because the particle and the hole occupy the same site. See Appendix~\ref{wf} for the analysis of the exciton wave function.

The situation is qualitatively different for a metallic  noninteracting system {\it with  perfect FS nesting}. If ${\text{min}}(b_{{\bm k}})=0$, the condition  \eqref{eq:pole_L} cannot be fulfilled because $\Phi(\omega<\omega_{ph}) \leq \Phi(\omega_{ph}) = 1/U$. Therefore, the LM of the AFM phase is absent in this case.

\section{Longitudinal mode of bilayer iridate}
From the above analysis, we predict that the  bilayer iridate Sr$_3$Ir$_2$O$_7$  should exhibit a LM   in a relatively small region around the center of the Brillouin zone. This prediction is qualitatively different from a previous interpretation of the RIXS data~\cite{Moretti15}, based on a pure spin model (i.e. strong-coupling limit), that reports the existence of a LM over the whole Brillouin zone. The effective Hamiltonian~(\ref{eq:H_bl}) for Sr$_3$Ir$_2$O$_7$ can be obtained by constructing a tight-binding model of the $t_{2g}$ orbitals and projecting it onto $J_{\rm eff}=1/2$ lowest energy doublet~\cite{Carter13}. The resulting  parameters of the effective single-orbital Hubbard model can be optimized to reproduce the experimentally observed magnon dispersion~\cite{Kim12,Moretti15,Hogan16,Lu17}: $\alpha=1.45$ and $t=0.11, t_z=0.09, U=0.33$ in units of eV. This value of $\alpha \approx \pi/2$ produces a strong easy-axis anisotropy, realized in a tetragonal elongation of octahedra consistent with $|t|>|t_z|$, as estimated for Sr$_3$Ir$_2$O$_7$~\cite{Carter13}.

The dynamical spin structure factor
\begin{equation}
S^{\mu\nu}(\boldsymbol{q},\omega)=\int_{\infty}^{\infty}dte^{it\omega}\langle S_{\boldsymbol{q}}^{\mu}(t)S_{-\boldsymbol{q}}^{\nu}(0)\rangle,
\end{equation}
where $S_{\boldsymbol{q}}^{\mu}=\frac{1}{\sqrt{N}}\sum_{{\bm r}}S_{\bm r}^{\mu}e^{i\boldsymbol{q}\cdot\boldsymbol{r}}$, is obtained from the dynamical spin susceptibility given in Eqs.~\eqref{eq:Trans_pm} and \eqref{eq:rpa_L}. Figure~\ref{fig:sqo} shows the out-of-phase ($q_z=\pi$) transverse (OT) response, $S^{xx}(\boldsymbol{q},\omega)=S^{yy}(\boldsymbol{q},\omega)$, and the out-of-phase longitudinal (OL) response, $S^{zz}(\boldsymbol{q},\omega)$, for $U=0.33$, $U_{\rm c} \approx 0.28$, and $0.23$~eV. The in-phase, or $q_z=0$, transverse (IT) response is not shown because it is practically identical to the OT response~\cite{Kim12}. This is a direct consequence of the strong easy-axis effective interlayer exchange that suppresses the single magnon tunneling between the two layers. In the large-$U$ limit, the effective interlayer exchange includes only Ising and Dzyaloshinskii-Moriya exchange interactions that do not split the two degenerate single-layer modes. This anisotropy is also responsible for the large spin gap revealed by RIXS measurements~\cite{Kim12}. No excitation peak is found below the particle-hole continuum in the in-phase longitudinal response. To match the RIXS peak at $(\frac{\pi}{2},\frac{\pi}{2})$, we added the intralayer next nearest neighbor hopping $0.012$~eV $\approx t/10$, which only changes the eigenenergy in the above argument. The overall dispersion curve for $U=0.33$~eV, shown in Figs.~\ref{fig:sqo}(a) and \ref{fig:sqo}(b), is consistent with the RIXS measurements~\cite{Kim12,Moretti15,Hogan16,Lu17}. The resultant charge gap~($\approx 130$~meV) is also consistent with the experimental observation~\cite{Okada13}. The structure of the continuum reflects the underlying electron bands. Unfortunately, the resolution of the reported RIXS data~\cite{Kim12,Moretti15,Hogan16,Lu17} is not enough to extract structures in the continuum. It would be of interest to compare the structure and the onset of the continuum to our calculation.

The most salient feature of the results shown in Fig.~\ref{fig:sqo}  is the  sharp OL mode near ${\bm q}_{\bot}=(0,0)$ and $(\pi,\pi)$.
Interestingly, the sharp OL mode appears only at restricted wave vectors because the particle-hole continuum exists in the same energy scale. The OL mode becomes gapless at $U=U_{\rm c}$, while the OT mode remains gapped, as shown in Figs.~\ref{fig:sqo}(c) and \ref{fig:sqo}(d); the exciton peaks in the band insulator for $U < U_{\rm c}$ remain sharp, as shown in Figs.~\ref{fig:sqo}(e) and \ref{fig:sqo}(f).
The $U$ dependence of the energy of this OL mode has important consequences for its stability.
While the kinematic constraints do not allow for its decay into two transverse modes for the parameters of  Sr$_{3}$Ir$_{2}$O$_{7}$, the fact that $\omega_L$ ($\omega_T$) is of order  $U$ ($t^2/U$) in the large-$U/t$ limit implies that the decay becomes kinematically allowed above a certain value of $U$. In other words, the sharp OL mode that we predict for Sr$_{3}$Ir$_{2}$O$_{7}$ is protected by the  large spin gap generated by the easy-axis anisotropy {\it and by the relatively small value of $U/t$}. Thus, the sharpness of the OL peak and the fact that $\omega_L$ and $\omega_T$ are comparable energy scales are clear indicators of the proximity of  Sr$_{3}$Ir$_{2}$O$_{7}$ to the QCP at $U=U_{\rm c}$. Because the real material has a small inter-bilayer hopping, the dimension of the effective theory that describes this QCP, $D=3+1$, coincides with the upper critical dimension (Gaussian fixed point). This means that  the mean field theory adopted here  is correct  up to logarithmic corrections.

Finally, we note that  higher order processes through the $J_{\rm eff}=3/2$ orbitals induce further neighbor hopping terms that can make the noninteracting system semimetallic without the FS nesting. In this case, a first-order metal-to-insulator transition occurs at a finite value of $U$~\cite{Carter13}, implying that there are two alternative scenarios for suppressing the AFM order via reduction of the coupling strength ($U/t$) in bilayer iridates. For the case of Sr$_{3}$Ir$_{2}$O$_{7}$, we predict that, if the material transitions into a band insulator, a QCP must exist at $U_{\rm c} \approx 0.28$~eV, which is only 15\% smaller than the estimated value  ($U=0.33$~eV) at ambient pressure. The coupling strength $U/t$ can be reduced by applying high pressure~\cite{Zhang19}. In this scenario, the LM becomes soft at 
$U=U_c$. In contrast, if the transition is of first order, the material becomes metallic without the softening of the LM. While absent in a metallic phase, the characteristic $S^z=0$ exciton peak appears in the bilayer spin-orbit band insulator for $U < U_{\rm c}$, as shown in Fig.~\ref{fig:sqo}(f).

\section{Summary and discussion}
We have reported the existence  of a sharp LM in antiferromagnetic bilayer Mott insulators with large spin-orbit coupling. 
Whenever the noninteracting system is a band insulator, $S^z=0,\pm 1$ excitons emerge from the particle-hole continuum for an infinitesimally small value of $U$. For materials with easy-axis anisotropy, such as the bilayer iridate Sr$_{3}$Ir$_{2}$O$_{7}$,  the Ising-like AFM ordering results  from the condensation of  $S^z=0$ excitons at a critical value $U=U_{\rm c}$. If the noninteracting system is a metal with FS nesting, the formation and condensation of the particle-hole pairs that produce the magnetic moments occur {\it simultaneously} at $U_c=0^+$.
Similarly to the case of  single-layer Mott insulators, such as Sr$_{2}$IrO$_{4}$, the LM is absent in the resulting AFM state. 

We emphasize that the  LM that we discussed in this work exists as a sharp mode only close enough to the QCP 
between the AFM phase and the paramagnetic band insulator. A similar QCP still exists in the large-$U$ limit, where it divides the AFM state from a quantum paramagnetic spin state, whose mean field description is a product state of rung singlets~\cite{Moretti15}, which is adiabatically connected with the band insulator. Note that such a mean field state can only be captured by expanding the variational space of  {\it product single-site sates} that is assumed by the conventional AFM mean field approach used in this work. However, our estimate of the hopping amplitudes for Sr$_3$Ir$_2$O$_7$, consistent with the effective model obtained from the three-orbital model~\cite{Carter13}, leads to $J_z/J=(t_z/t)^2 \approx 0.67$ in the strong coupling limit. This value of the exchange coupling ratio is significantly smaller than the critical value  $(J_z/J)_{\rm c}$. The bond-operator mean field theory
gives $(J_z/J)_{\rm c}=4$ for the isotropic and the easy-axis cases~\cite{Su20,Moretti15}, while the exact value for the SU(2) invariant case is very close to $(J_z/J)_{\rm c}=2.5221$ according to quantum Monte Carlo simulations~\cite{Sen15}. Therefore, as we discussed in the previous section, the sharp LM mode does not survive in the large-$U$ limit of the Hubbard model of Sr$_3$Ir$_2$O$_7$ (the LM loses its sharpness far enough from the QCP because the kinematic constraints allow it to decay into pairs of transverse modes). Moreover, as shown in Fig.~\ref{fig:sqo}, the sharp LM appears only at restricted wave vectors because the particle-hole continuum exists in the same energy scale. This characteristic feature of Sr$_3$Ir$_2$O$_7$ indicates that this material is an excitonic insulator~\cite{Cloizeaux65,Halperin68,Jerome67} near the critical point $U=U_c$. In other words, the proximity of  Sr$_3$Ir$_2$O$_7$ to quantum criticality is caused by strong charge fluctuations, which are absent in the pure spin model.

Finally, it is worth mentioning that the longitudinal mode can also exist  in bilayer Mott insulators with easy-plane anisotropy.
The corresponding $XY$-AFM state results from the condensation of the $S^z=\pm 1$ excitons, while the $S^z=0$ exciton remains gapped at 
$U=U_c$ and becomes the LM for $U>U_c$. The main difference is that this LM, also known as ``Higgs mode,'' is critically damped in (3+1)D because it is allowed to decay into two transverse magnons (Goldstone modes) of the AFM state~\cite{Affleck92,Kulik11,Podolsky11,Podolsky12,Lohoefer15,Qin17}. It is of great interest to seek the sharp LM and the Higgs mode in other physical systems.
Recently, fermionic systems described by bilayer Hubbard models have been realized in cold atoms~\cite{Gall21}, implying that the exciton condensation that we have discussed  here can also be realized in these systems.

\begin{acknowledgments}
  We thank Jian Liu and Mark Dean for helpful discussions. H.\nobreak\,S. acknowledges support from JSPS KAKENHI Grant No. JP19K14650.
S.-S.\nobreak\,Z. and C.\nobreak\,D.\nobreak\,B. are supported by funding from the Lincoln Chair of Excellence in Physics.
This research used resources of the Oak Ridge Leadership Computing
Facility at the Oak Ridge National Laboratory, which is supported
by the Office of Science of the U.S. Department of Energy under Contract
No. DE-AC05-00OR22725.
\end{acknowledgments}

\appendix

\section{Random phase approximation (RPA)}
\label{rpa}
In view of  the $U(1)$ invariance of ${\cal H}$  and of the ground state, the transverse and longitudinal  spin fluctuations
are decoupled from each other. The same is true for the charge fluctuations. Consequently, we rewrite the (bare) interaction vertex in three different forms which  account for the longitudinal and transverse spin fluctuations, respectively,
\begin{eqnarray}
{\cal H}_{int} & = & \frac{1}{2}\sum_{\bm{r}}\sum_{ \{\sigma_i\}}V_{\sigma_1 \sigma_4;\sigma_2\sigma_3}c_{\sigma_1,\bm{r}}^{\dagger} c_{\sigma_2,\bm{r}}^{\dagger}c_{\sigma_3,\bm{r}}c_{\sigma_4,\bm{r}},
\end{eqnarray}
where the interaction vertex takes three equivalent forms,
\begin{eqnarray}\label{eq:bare_vtx}
  V_{\sigma_1\sigma_4;\sigma_2\sigma_3} &=& {1\over	2}U \left( \sigma_{\sigma_1\sigma_4}^{0}\sigma_{\sigma_2\sigma_3}^{0} -\sigma_{\sigma_1\sigma_4}^{z}\sigma_{\sigma_2\sigma_3}^{z}\right) \\
  &=& {1\over	2}U \sigma_{\sigma_1\sigma_4}^{0}\sigma_{\sigma_2\sigma_3}^{0} - U \sigma_{\sigma_1\sigma_4}^{+}\sigma_{\sigma_2\sigma_3}^{-} \\
&=&{1\over 2}U \sigma_{\sigma_1\sigma_4}^{0}\sigma_{\sigma_2\sigma_3}^{0}-U \sigma_{\sigma_1\sigma_4}^{-}\sigma_{\sigma_2\sigma_3}^{+},
\end{eqnarray}
$\sigma^{0}$ is the $2\times 2$ identity matrix, and $\sigma^{\pm}=\frac{1}{2}(\sigma^{x}\pm i\sigma^{y})$.

\begin{figure}[tb]
\centering
\includegraphics[width=1.0\columnwidth]{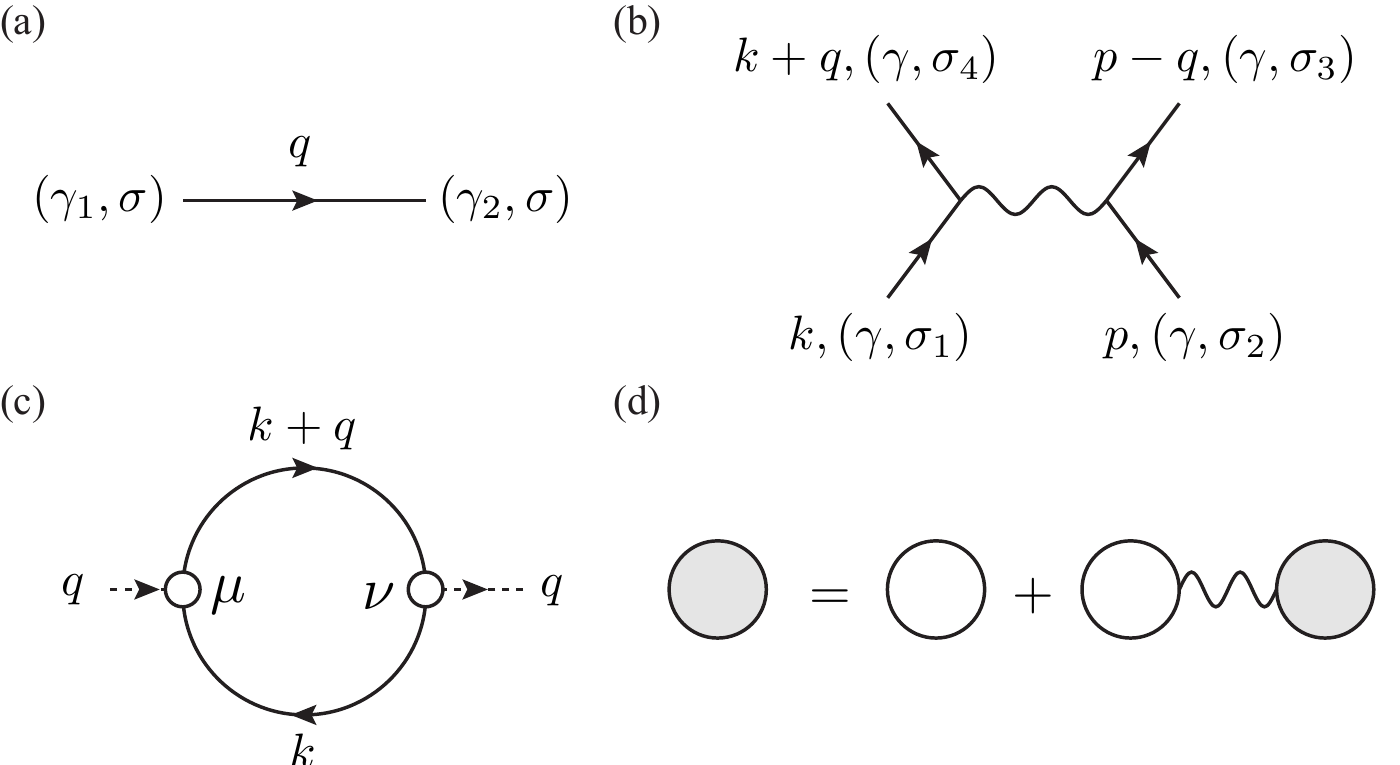}\caption{ {(a) Free fermion propagator in the background of the magnetic order: $-G_{0}(q)$, where $q\equiv({\bm q},i\omega_n)$. (b) Bare interaction vertex $-\frac{1}{{\cal N}_{u}\beta} V_{\sigma_1\sigma_4;\sigma_2\sigma_3} $.
(c) Bare and (d) RPA {spin-charge} susceptibility.} }
\label{fig:feyn}
\end{figure}
We are ready to compute the dynamic spin-charge susceptibility within the RPA.
Figures~\ref{fig:feyn}(a) and \ref{fig:feyn}(b) show  the fermion propagator and the bare vertex, respectively. The free fermion Green's function 
\begin{eqnarray}
G_{0}(\bm{q},i\omega_{n}) & = & \sum_{n=\pm,\sigma=\uparrow,\downarrow}\frac{P_{n\sigma}(\bm{q})}{i\omega_{n}-\varepsilon_{n\sigma}(\bm{q})}
\end{eqnarray}
is represented  by a solid line,  where $P_{n\sigma}(\bm{q})=\rvert X_{n\sigma}(\bm{q})\rangle\langle X_{n\sigma}(\bm{q})\rvert$ is the projector on the eigenstate $\rvert X_{n\sigma}(\bm{q})\rangle$ of the mean field
Hamiltonian.

The (bare) dynamic spin-charge susceptibility of the noninteracting mean field Hamiltonian is given by
\begin{align}
  &\chi_{0}^{(\gamma_1,\mu);(\gamma_2,\nu)}(\bm{q},i\omega_{n}) \\
  =& \frac{1}{{\cal N}_{u}}\sum_{\bm{k}} \sum_{k,l} \frac{n_{F}(\varepsilon_{l\sigma^{\prime}}(\bm{k}+\bm{q}))-n_{F}(\varepsilon_{k\sigma}(\bm{k}))}{i\omega_{n}-\left(\varepsilon_{l\sigma^{\prime}}(\bm{k}+\bm{q})-\varepsilon_{k\sigma}(\bm{k})\right)}F_{k\sigma;l\sigma^{\prime}}^{(\gamma_1,\mu);(\gamma_2,\nu)}, \nonumber
\end{align}
where $\gamma_1,\gamma_2={\cal A},{\cal B}$ are sublattice indices and $\mu,\nu=0,\pm, z$ are charge-spin indices. Here we have  introduced the polarization factor
\begin{align}
&\!\!\!\! F_{k\sigma;l\sigma^{\prime}}^{(\gamma_1,\mu);(\gamma_2,\nu)}(\bm{k};\bm{q}) = \text{Tr}\Bigg[P_{k\sigma}(\bm{k})\Sigma_{\gamma_1}^{\mu}P_{l\sigma^{\prime}}(\bm{k}+\bm{q})\Sigma_{\gamma_2}^{\nu}\Bigg] \\
 &\!\!\!\! = \langle X_{k\sigma}(\bm{k})\rvert\Sigma_{\gamma_1}^{\mu}\rvert X_{l\sigma^{\prime}}(\bm{k}+\bm{q})\rangle\langle X_{l\sigma^{\prime}}(\bm{k}+\bm{q})\rvert\Sigma_{\gamma_2}^{\nu}\rvert X_{k\sigma}(\bm{k})\rangle, \nonumber
\end{align}
with $\Sigma_{\gamma}^{\mu} \equiv {1\over 2}(\tau_0+\epsilon_\gamma \tau_z) \otimes \sigma^{\mu}$, where $\epsilon_\gamma$ takes the values 1 ($-1$) for $\gamma = {\mathcal A} (\mathcal B)$,
and $\tau_0$ and $\tau_z$ are the $2\times 2$ identity and the Pauli matrix of the sublattice space, respectively. The diagram of $\chi_{0}$ is shown in Fig.~\ref{fig:feyn}(c).
Because of the $U(1)$ spin rotation symmetry about the $z$ axis, $F_{k\sigma;l\sigma^{\prime}}^{(\gamma_1,\mu);(\gamma_2,\nu)}(\bm{k};\bm{q})$ has the following structure:
\begin{align}
  &F^{(\gamma_1,\mu);(\gamma_2,\nu)}\\
  = & \left(\begin{array}{cccc}
\!\!F^{(\gamma_1,0);(\gamma_2,0)} & 0 & 0 & 0\\ 
0& 0 & \!\!\!\!F^{(\gamma_1,+);(\gamma_2,-)} & 0\\
0& \!\!\!\!F^{(\gamma_1,-);(\gamma_2,+)} & 0 & 0\\
0& 0 & 0 & \!\!\!\!F^{(\gamma_1,z);(\gamma_2,z)}
\end{array}\!\! \right). \nonumber
\end{align}
The same structure holds for $\chi_{0}^{(\gamma_1,\mu);(\gamma_2,\nu)}$, confirming that the transverse
spin fluctuations, longitudinal spin fluctuations, and charge fluctuations are decoupled from each other. 
In the following, we focus on the spin channel, which is of main interest for this work.

Figure \ref{fig:feyn}(d) represents the magnetic susceptibility at the RPA level. The results are
\begin{eqnarray}
\!\!\!\! \!\!\!\! \!\!\!\!  \chi_{\rm RPA}^{+-}(\bm{q},i\omega_{n}) & = & \frac{1}{ \tau^{0}-U\chi_{0}^{+-}(\bm{q},i\omega_{n})}\chi_{0}^{+-}(\bm{q},i\omega_{n}), \\
\!\!\!\! \!\!\!\! \!\!\!\!  \chi_{\rm RPA}^{-+}(\bm{q},i\omega_{n}) & = & \frac{1}{\tau^{0}-U\chi_{0}^{-+}(\bm{q},i\omega_{n})}\chi_{0}^{-+}(\bm{q},i\omega_{n})
\end{eqnarray}
for the transverse channel and
\begin{eqnarray}
\!\!\!\! \!\!\!\! \!\!\!\! \chi_{\rm RPA}^{zz}(\bm{q},i\omega_{n}) & = & \frac{1}{\tau^{0}-{U\over 2}\chi_{0}^{zz}(\bm{q},i\omega_{n})}\chi_{0}^{zz}(\bm{q},i\omega_{n})
\end{eqnarray}
for the longitudinal channel.
Note that $\chi_{\rm RPA}^{+-}$, $\chi_{\rm RPA}^{-+}$, and $\chi_{\rm RPA}^{zz}$ are $2\times2$ matrices in the sublattice space, while $\chi_{0}^{+-}$, $\chi_{0}^{-+}$, and $\chi_{0}^{zz}$ are the corresponding bare magnetic susceptibilities, respectively.

\section{Exciton wave function}
\label{wf}
We here investigate the wave function of the exciton formed by multiple particle-hole pairs with $S^{z}=0$. In the large-$U$ limit,
it is mainly composed of a single particle-hole pair because the energy cost of creating a 
particle-hole is of order $\sim U$. The wave function of the pair is anticipated
to be a tightly bound state because the
mean field bandwidth of an electron/hole $\sim t^{2}/U$ is comparable
to the binding energy $\zeta t_z^{2}/U$, where $\zeta\simeq 0.88$ obtained
by solving the pole equation for the relevant set of the hopping parameters.

To determine the size of the bound state, we
consider the particle-hole Green's function
\begin{align}
  &i{\cal G}_{(\gamma_{1}\sigma_{1},\gamma_{1}^{\prime}\sigma_{1}^{\prime});(\gamma_{2}\sigma_{2},\gamma_{2}^{\prime}\sigma_{2}^{\prime})}^{(2)}(\bm{Q},\omega;\bm{k},\bm{q}) \nonumber \\
  = & \int_{-\infty}^{\infty}dte^{i\omega t}\langle G\rvert T_{t}c_{\gamma_{1}\sigma_{1}}^{\dagger}(\bm{k}-\frac{\bm{Q}}{2},t)c_{\gamma_{1}^{\prime}\sigma_{1}^{\prime}}(\bm{k}+\frac{\bm{Q}}{2},t)\nonumber \\
 & \qquad \quad \times c_{\gamma_{2}^{\prime}\sigma_{2}^{\prime}}^{\dagger}(\bm{q}+\frac{\bm{Q}}{2},0)c_{\gamma_{2}\sigma_{2}}(\bm{q}-\frac{\bm{Q}}{2},0)\rvert G\rangle.
\end{align}
The exciton gives rise to a pole at $\omega=\omega_L({\bm Q})$ for the center-of-mass momentum ${\bm Q}$:
\begin{align}
  & i{\cal G}_{(\gamma_{1}\sigma_{1},\gamma_{1}^{\prime}\sigma_{1}^{\prime});(\gamma_{2}\sigma_{2},\gamma_{2}^{\prime}\sigma_{2}^{\prime})}^{(2)}(\bm{Q},\omega;\bm{k},\bm{q}) \nonumber \\
  \sim & \frac{\psi_{\gamma_{1}\sigma_{1};\gamma_{1}^{\prime}\sigma_{1}^{\prime}}(\bm{k})\psi_{\gamma_{2}\sigma_{2},\gamma_{2}^{\prime}\sigma_{2}^{\prime}}^{*}(\bm{q})}{\omega-\omega_L({\bm Q}) +i0^{+}}+\text{regular term},
\end{align}
where $\psi_{\gamma_{1}\sigma_{1};\gamma_{1}^{\prime}\sigma_{1}^{\prime}}(\bm{k})=\langle G\rvert c_{\gamma_{1}\sigma_{1}}^{\dagger}(\bm{k}-\frac{\bm{Q}}{2})c_{\gamma_{1}^{\prime}\sigma_{1}^{\prime}}(\bm{k}+\frac{\bm{Q}}{2})  \rvert b_{\bm Q}\rangle  $, $\rvert b_{\bm Q}\rangle$ is the exciton eigenstate with center of mass momentum ${\bm Q}$, and $\omega_L({\bm Q}) $ is
the exciton energy measured from the ground state.  The probability amplitude to find a
hole with spin $\sigma_{1}$ at $\bm{r}_{1}$ and an electron with spin
$\sigma_{2}$ at $\bm{r}_{2}$ 
is obtained by Fourier transforming  the wave function $\psi_{\gamma_{1}\sigma_{1};\gamma_{1}^{\prime}\sigma_{1}^{\prime}}(\bm{k})$:
\begin{align}
  & \psi_{\sigma_{1};\sigma_{2}}(\bm{r}_{1}\in\gamma_{1},\bm{r}_{2}\in\gamma_{2}) = \langle G\rvert c_{\sigma_{1}}^{\dagger}(\bm{r}_{1})c_{\sigma_{2}}(\bm{r}_{2})\rvert b\rangle\nonumber \\
 = & e^{i\bm{Q}\cdot(\bm{r}_{1}+\bm{r}_{2})/2}\frac{1}{{\cal N}_{u}}\sum_{\bm{k}}\psi_{\gamma_{1}\sigma_{1};\gamma_{2}\sigma_{2}}(\bm{k})e^{i\bm{k}\cdot(\bm{r}_{2}-\bm{r}_{1})}.
\end{align}

\begin{figure}[t]
\centering
\includegraphics[width=1.0\columnwidth]{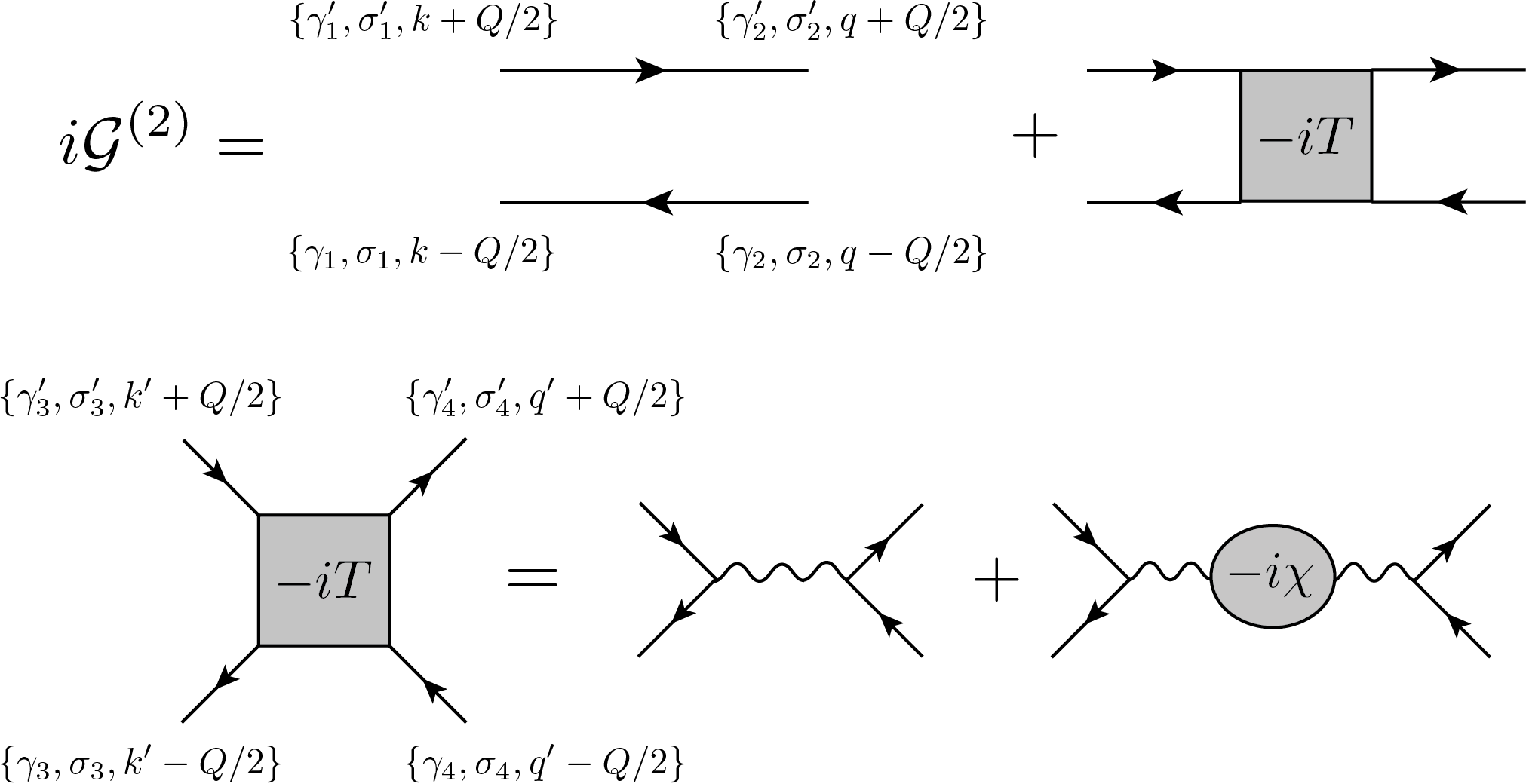}\caption{The particle-hole Green's function and the $T$ matrix under the RPA.}
\label{fig:phgreen}
\end{figure}
As shown in Fig.~\ref{fig:phgreen}, the particle-hole Green's function is determined by the $T$ matrix:
\begin{align}
  & i{\cal G}_{(\gamma_{1}\sigma_{1},\gamma_{1}^{\prime}\sigma_{1}^{\prime});(\gamma_{2}\sigma_{2},\gamma_{2}^{\prime}\sigma_{2}^{\prime})}^{(2)}(\bm{Q},\omega;\bm{k},\bm{q}) \nonumber \\
  = & iG_{(\gamma_{1}\sigma_{1},\gamma_{1}^{\prime}\sigma_{1}^{\prime});(\gamma_{2}\sigma_{2},\gamma_{2}^{\prime}\sigma_{2}^{\prime})}^{(2)}(\bm{Q},\omega;\bm{k},\bm{q}) \nonumber \\
  & +iG_{(\gamma_{1}\sigma_{1},\gamma_{1}^{\prime}\sigma_{1}^{\prime});(\gamma_{3}\sigma_{3},\gamma_{3}^{\prime}\sigma_{3}^{\prime})}^{(2)}(\bm{Q},\omega;\bm{k},\bm{k}^{\prime}) \\
  & \quad (-i)T_{(\gamma_{3}\sigma_{3},\gamma_{3}^{\prime}\sigma_{3}^{\prime});(\gamma_{4}\sigma_{4},\gamma_{4}^{\prime}\sigma_{4}^{\prime})}(\bm{Q},\omega;\bm{k}^{\prime},\bm{q}^{\prime}) \nonumber \\
& \quad iG_{(\gamma_{4}\sigma_{4},\gamma_{4}^{\prime}\sigma_{4}^{\prime});(\gamma_{2}\sigma_{2},\gamma_{2}^{\prime}\sigma_{2}^{\prime})}^{(2)}(\bm{Q},\omega;\bm{q}^{\prime},\bm{q}). \nonumber
\end{align}
Here $iG^{(2)}$ is the noninteracting particle-hole Green's function
\begin{widetext}
  \begin{align}
 &  iG_{(\gamma_{1}\sigma_{1},\gamma_{1}^{\prime}\sigma_{1}^{\prime});(\gamma_{2}\sigma_{2},\gamma_{2}^{\prime}\sigma_{2}^{\prime})}^{(2)}(\bm{Q},\omega;\bm{k},\bm{q})\nonumber \\
 = & \int_{-\infty}^{\infty}dte^{i\omega t}\langle G\rvert T_{t}c_{\gamma_{1}\sigma_{1}}^{\dagger}(\bm{k}-\frac{\bm{Q}}{2},t)c_{\gamma_{1}^{\prime}\sigma_{1}^{\prime}}(\bm{k}+\frac{\bm{Q}}{2},t)c_{\gamma_{2}^{\prime}\sigma_{2}^{\prime}}^{\dagger}(\bm{q}+\frac{\bm{Q}}{2},0)c_{\gamma_{2}\sigma_{2}}(\bm{q}-\frac{\bm{Q}}{2},0)\rvert G\rangle  \\
 = & \delta_{\bm{k},\bm{q}}\int_{-\infty}^{\infty}dte^{i\omega t}G_{\gamma_{1}^{\prime}\sigma_{1}^{\prime};\gamma_{2}^{\prime}\sigma_{2}^{\prime}}(\bm{k}+\frac{\bm{Q}}{2},t)G_{\gamma_{2}\sigma_{2};\gamma_{1}\sigma_{1}}(\bm{k}-\frac{\bm{Q}}{2},-t)  \\
 = & \delta_{\bm{k},\bm{q}}\int_{-\infty}^{\infty}\frac{d\nu}{2\pi}G_{\gamma_{1}^{\prime}\sigma_{1}^{\prime};\gamma_{2}^{\prime}\sigma_{2}^{\prime}}(\bm{k}+\frac{\bm{Q}}{2},\nu+\frac{\omega}{2})G_{\gamma_{2}\sigma_{2};\gamma_{1}\sigma_{1}}(\bm{k}-\frac{\bm{Q}}{2},\nu-\frac{\omega}{2})  \\
 = & i\delta_{\bm{k},\bm{q}}\delta_{\sigma_{1}\sigma_{2}}\delta_{\sigma_{1}^{\prime}\sigma_{2}^{\prime}}\Bigg[\frac{\left(P_{1,\sigma_{1}^{\prime}}(\bm{k}+\frac{\bm{Q}}{2})\right)_{\gamma_{1}^{\prime}\sigma_{1}^{\prime};\gamma_{2}^{\prime}\sigma_{2}^{\prime}}\left(P_{-1,\sigma_{1}}(\bm{k}-\frac{\bm{Q}}{2})\right)_{\gamma_{2}\sigma_{2};\gamma_{1}\sigma_{1}}}{\omega-(\varepsilon_{1,\bm{k}+\frac{\bm{Q}}{2}}-\varepsilon_{-1,\bm{k}-\frac{\bm{Q}}{2}})}  \\
 & -\frac{\left(P_{-1,\sigma_{1}^{\prime}}(\bm{k}+\frac{\bm{Q}}{2})\right)_{\gamma_{1}^{\prime}\sigma_{1}^{\prime};\gamma_{2}^{\prime}\sigma_{2}^{\prime}}\left(P_{1,\sigma_{1}}(\bm{k}-\frac{\bm{Q}}{2})\right)_{\gamma_{2}\sigma_{2};\gamma_{1}\sigma_{1}}}{\omega+\varepsilon_{1,\bm{k}-\frac{\bm{Q}}{2}}-\varepsilon_{-1,\bm{k}+\frac{\bm{Q}}{2}}}\Bigg], \nonumber
  \end{align}
\end{widetext}
where $P_{s,\sigma}(\bm{k})=\rvert X_{s,\sigma}(\bm{k})\rangle\langle X_{s,\sigma}(\bm{k})\rvert$
is the projector to the $(s,\sigma)$ eigenstate at $\bm{k}$. In the RPA (see Fig.~\ref{fig:phgreen}), the $T$ matrix is given by
\begin{align}
  & -iT_{(\gamma_{3}\sigma_{3},\gamma_{3}^{\prime}\sigma_{3}^{\prime});(\gamma_{4}\sigma_{4},\gamma_{4}^{\prime}\sigma_{4}^{\prime})}(\bm{Q},\omega;\bm{k}^{\prime},\bm{q}^{\prime}) \nonumber \\
  = & \ i\frac{U}{2}\sigma^{z}_{\sigma_{3}\sigma_{3}^{\prime}}\sigma^{z}_{\sigma_{4}\sigma_{4}^{\prime}}\delta_{\gamma_{3}\gamma_{3}^{\prime}}\delta_{\gamma_{4}\gamma_{4}^{\prime}}\delta_{\gamma_{3}\gamma_{4}} \label{Tmatrix} \\
  & + i\frac{U}{2}\sigma^{z}_{\sigma_{3}\sigma_{3}^{\prime}}\delta_{\gamma_{3}\gamma_{3}^{\prime}}(-i)\chi_{\rm RPA}^{(\gamma_{3}z,\gamma_{4}z)}(\bm{Q},\omega)i\frac{U}{2}\sigma^{z}_{\sigma_{4}\sigma_{4}^{\prime}}\delta_{\gamma_{4}\gamma_{4}^{\prime}}. \nonumber
\end{align}
Note that the bare Hubbard interaction is written in terms of the longitudinal  component of the spin operator, 
${\cal H}_{int} = \sum_{\bm{r}} [(1/4) (n_{\bm r})^2 - (S^z_{\bm r})^2]$ with $n_{\bm r}=\sum_{\sigma}c_{\sigma}^{\dagger}(\bm r)c_{\sigma}(\bm r)$ and $S^z_{\bm r}=(1/2)\sum_{\sigma}\sigma c_{\sigma}({\bm r})^{\dagger}c_{\sigma}({\bm r})$. Under the RPA, the $T$ matrix that describes the renormalized interaction between electrons includes contributions from longitudinal spin fluctuations. 

\begin{figure*}
\centering
\includegraphics[width=2.0\columnwidth]{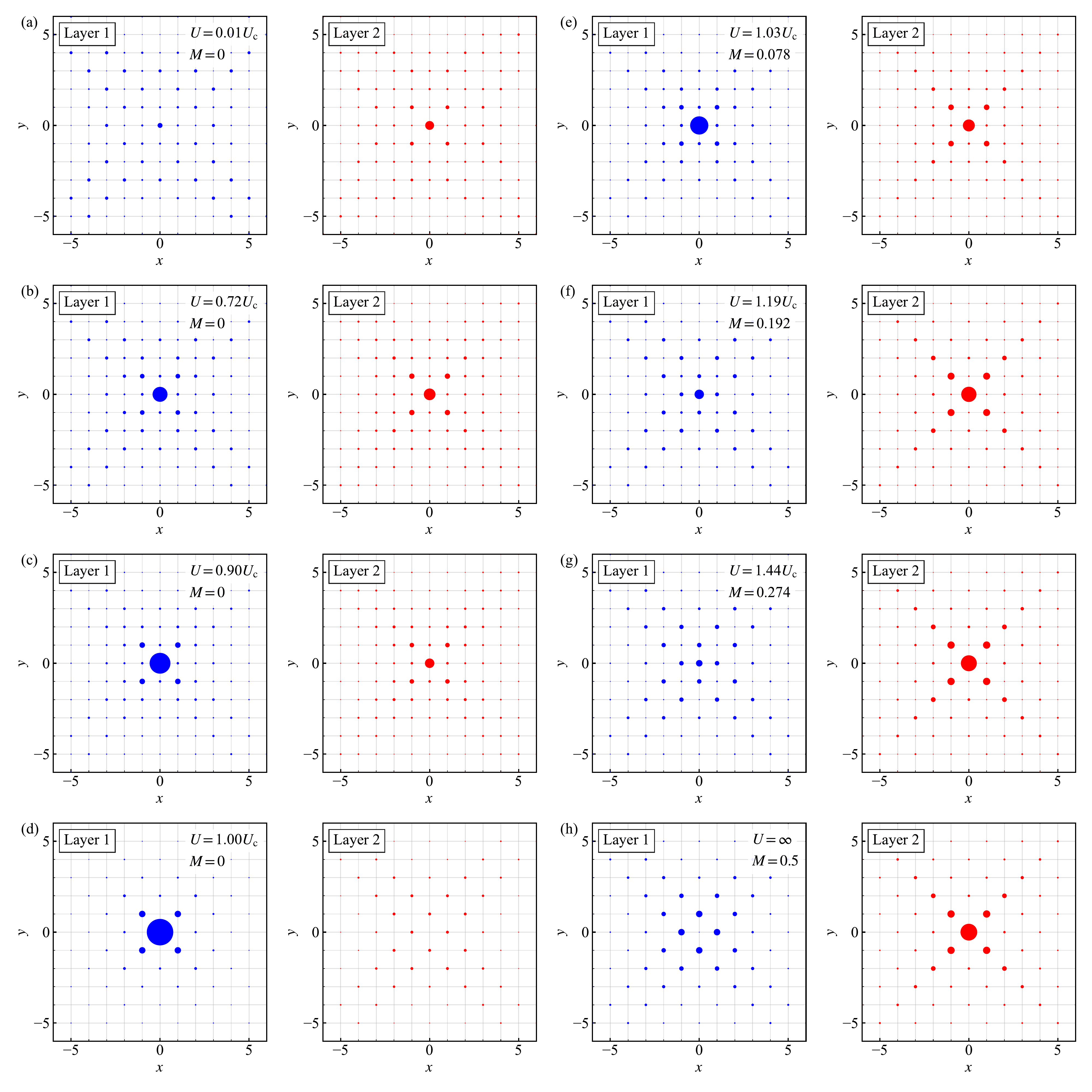}\caption{Probability distribution $\rvert\psi_{\downarrow\downarrow}(\bm{0},\bm{r}_{2})\rvert^{2}$ for several strengths of the on-site coupling $U$. A hole with spin $\downarrow$ is assumed located at $\bm{r}_1=\bm{0}$ on layer 1. The critical coupling is $U_{\rm c} \approx 0.277$~eV for the hopping parameters relevant to Sr$_3$Ir$_2$O$_7$, while the 
on-site Hubbard interaction   is estimated to be $U=0.33$ eV $(=1.19 U_{\rm c})$ for the material. $M$ is the magnitude of the local ordered magnetic moment. The distribution function $\rvert\psi_{\downarrow\downarrow}(\bm{0},\bm{r}_{2})\rvert^{2}$
is normalized for each value of $U$,  so that the sum over $\bm{r}_{2}$ is equal to one.}
\label{wf:evolution}
\end{figure*}
Let us consider the exciton at $\bm{Q}=\bm{0}$, which has the lowest energy $\omega_L({\bm Q} = {\bm 0}) \equiv \omega_L$. Near the exciton pole, the longitudinal RPA susceptibility is given by
\begin{eqnarray}
\!\!\!\! \chi_{\rm RPA}^{(\gamma_{2}z,\gamma_{2}^{\prime}z)}(\bm{0},\omega) & \sim & -\frac{(-1)^{\gamma_{2}+\gamma_{2}^{\prime}}}{U^{2}\Phi^{\prime}(\omega_L)}\frac{1}{\omega-\omega_L+i0^{+}},
\end{eqnarray}
where
\begin{eqnarray}
\!\!\!\!\!\!\!\!  \Phi^{\prime}(\omega_L) & = & \frac{\omega_L}{4(UM)^{3}}\frac{1}{{\cal N}_{u}}\sum_{\bm{k}\in BZ}\frac{z_{\bm{k}}^{3}(1-z_{\bm{k}}^{2})}{\left[1-(\frac{\omega_L}{2UM})^{2}z_{\bm{k}}^{2}\right]^{2}}  
\end{eqnarray}
for $U>U_{\rm c}$.
In the vicinity of the critical point $U_{\rm c}$, it takes the asymptotic form $\Phi^{\prime}(\omega_L)\simeq UM\frac{1}{{\cal N}_{u}}\sum_{\bm{k}}\rvert b_{\bm{k}}\rvert^{-3}\propto\sqrt{U-U_{\rm c}}$. The longitudinal RPA susceptibility, therefore, acquires a diverging
spectral weight $\propto1/\sqrt{U-U_{\rm c}}$, implying strong longitudinal
spin fluctuations. In the large-$U$ limit,
\begin{equation}
\Phi^{\prime}(\omega_L)\rightarrow\frac{1}{{\cal N}_{u}}\sum_{\bm{k}}\frac{2b_{\bm{k}}^{2}}{(\zeta t^{2}+2b_{\bm{k}}^{2})^{2}}\sim{\cal O}(1),
\end{equation}
and the spectral weight of the longitudinal RPA susceptibility approaches zero as $U^{-2}$.

According to Eq.~(\ref{Tmatrix}), the exciton mode also appears as a pole of the particle-hole Green's function. For instance,
\begin{align}
  & {\cal G}_{(\gamma_{1}\sigma_{1},\gamma_{1}^{\prime}\sigma_{1}^{\prime});(\gamma_{2}\sigma_{2},\gamma_{2}^{\prime}\sigma_{2}^{\prime})}^{(2)}(\bm{0},\omega;\bm{k},\bm{q}) \nonumber \\
  \sim & \frac{1}{\omega-\omega_L+i0^{+}}\frac{1}{\Phi^{\prime}(\omega_L)}\sigma_{1}\sigma_{2}\delta_{\sigma_{1}\sigma_{1}^{\prime}}\delta_{\sigma_{2}\sigma_{2}^{\prime}}  \nonumber \\
 & \times \sum_{\gamma_{3}}(-1)^{\gamma_{3}}G_{(\gamma_{1}\sigma_{1},\gamma_{1}^{\prime}\sigma_{1});(\gamma_{3}\sigma_{1},\gamma_{3}\sigma_{1})}^{(2)}(\bm{0},\omega_L;\bm{k},\bm{k})\\
  & \times \sum_{\gamma_{4}}(-1)^{\gamma_{4}}G_{(\gamma_{4}\sigma_{2},\gamma_{4}\sigma_{2});(\gamma_{2}\sigma_{2},\gamma_{2}^{\prime}\sigma_{2})}^{(2)}(\bm{0},\omega_L;\bm{q},\bm{q})  \nonumber \\
  & +\text{regular terms}.  \nonumber
\end{align}
The spectral weight of the exciton pole is equal to the projection
of the exciton wave function to the two-magnon sector
\begin{align}
  & \psi_{\gamma\sigma;\gamma^{\prime}\sigma^{\prime}}(\bm{k}) \\
  =& \frac{\sigma}{\sqrt{\Phi^{\prime}(\omega_L)}}\delta_{\sigma\sigma^{\prime}}\sum_{\gamma_{3}}(-1)^{\gamma_{3}}G_{(\gamma\sigma;\gamma^{\prime}\sigma);(\gamma_{3}\sigma,\gamma_{3}\sigma)}^{(2)}(\bm{0},\omega_L;\bm{k},\bm{k}). \nonumber
\end{align}
For each combination of sublattices, it reads
\begin{align}
  \psi_{\mathcal{A}\sigma;\mathcal{A}\sigma^{\prime}}(\bm{k}) & = \frac{\sigma}{2\sqrt{\Phi^{\prime}(\omega_L)}}\delta_{\sigma\sigma^{\prime}}(1-z_{\bm{k}}^{2}) \\
  & \qquad \times \frac{\varepsilon_{1,\bm{k}}-\varepsilon_{-1,\bm{k}}}{\omega_L^{2}-(\varepsilon_{1,\bm{k}}-\varepsilon_{-1,\bm{k}})^{2}}, \nonumber \\
\psi_{\mathcal{B}\sigma;\mathcal{B}\sigma^{\prime}}(\bm{k}) & = -\psi_{\mathcal{A}\sigma;\mathcal{A}\sigma^{\prime}}(\bm{k}),\\
\psi_{\mathcal{A}\sigma;\mathcal{B}\sigma^{\prime}}(\bm{k}) & = \frac{\sigma}{\sqrt{\Phi^{\prime}(\omega_L)}}\delta_{\sigma\sigma^{\prime}}x_{\sigma\bm{k}}^{*}\frac{\sqrt{1-z_{\bm{k}}^{2}}}{2} \\
& \times \left[\frac{1-\sigma z_{\bm{k}}}{\omega_L-(\varepsilon_{1,\bm{k}}-\varepsilon_{-1,\bm{k}})}+\frac{1+\sigma z_{\bm{k}}}{\omega_L+\varepsilon_{1,\bm{k}}-\varepsilon_{-1,\bm{k}}} \right], \nonumber \\
\psi_{\mathcal{B}\sigma;\mathcal{A}\sigma}(\bm{k}) & = \psi_{\mathcal{A}\bar{\sigma};\mathcal{B}\bar{\sigma}}^{*}(\bm{k}).
\end{align}

In the large-$U$ limit, $\psi_{\mathcal{A}\sigma;\mathcal{A}\sigma^{\prime}}(\bm{k})\sim{\cal O}(U^{-1})$,
$\psi_{\mathcal{A}\uparrow;\mathcal{B}\uparrow}(\bm{k})=\psi_{\mathcal{B}\downarrow;\mathcal{A}\downarrow}^{*}(\bm{k})\sim{\cal O}(U^{-2})$,
and $\psi_{\mathcal{A}\downarrow;\mathcal{B}\downarrow}(\bm{k})=\psi_{\mathcal{B}\uparrow;\mathcal{A}\uparrow}^{*}(\bm{k})\sim{\cal O}(1)$.
The asymptotic form of the nonzero amplitude reads
\begin{align}
  & \psi_{\mathcal{A}\downarrow;\mathcal{B}\downarrow}(\bm{k})=\psi_{\mathcal{B}\uparrow;\mathcal{A}\uparrow}^{*}(\bm{k}) \nonumber \\
  = & \frac{1}{\sqrt{\Phi^{\prime}(\omega_L)}}\frac{2b_{\bm{k}}^{*}}{\zeta t^{2}+2(b_{\bm{k}}^{2} - \min_{q} b_{\bm{q}}^2)},
\end{align}
where $\zeta\simeq 0.88$ determines the exciton binding energy $E_{b}\equiv \omega_{ph}-\omega_L \simeq \zeta t_z^{2}/U$. The amplitude $\psi_{\downarrow\downarrow}(\bm{r}_{1}\in \mathcal{A},\bm{r}_{2}\in \mathcal{B})$  determines the spectral weight of the configuration  with one  hole with spin $\downarrow$ at $\bm{r}_{1}\in \mathcal{A}$
and one electron with spin $\downarrow$ at $\bm{r}_{2}\in \mathcal{B}$.
Since  the ordered moment is $-M$ on sublattice $\mathcal{A}$ and $M$
on sublattice $\mathcal{B}$,  the exciton is created
by moving either $\downarrow$ spin from sublattice $\mathcal{A}$ to $\mathcal{B}$ or
a $\uparrow$ spin from sublattice $\mathcal{B}$ to $\mathcal{A}$. On the condition that
a hole with spin $\downarrow$ is pinned at $\bm{r}_{1}=\bm{0}$ (sublattice
$\mathcal{A}$ and layer 1), the distribution of the spectral weight, i.e., $\rvert\psi_{\downarrow\downarrow}(\bm{0},\bm{r}_{2})\rvert^{2}$, is shown in Fig.~\ref{wf:evolution}. In the large-$U$ limit, a hole and an electron occupy different sublattices [Fig.~\ref{wf:evolution}(h)]. The bound state with the particle-hole pair  occupying a vertical bond has the largest spectral weight and the probability of finding the particle and the hole on different layers is higher than the probability of finding them on the same layer. As a result of the  strong binding energy relative to
the bandwidth of the single electron or hole ($\sim E_{b}$), the size of the bound state is comparable to one lattice constant.

Figure~\ref{wf:evolution} shows the evolution of the real space distribution of the particle-hole pair as a function of $U$. The ordered magnetic moment, $M=|\langle G\rvert S_{\bm{r}}^{z}\rvert G\rangle|$ decreases
upon reducing $U/t$ within the ordered phase [Figs.~\ref{wf:evolution}(e)--\ref{wf:evolution}(g)] and the probability of finding the particle and the hole on the same site increases. Thus, $S_{\bm{r}}^{z}\rvert G\rangle$ provides a reasonably good approximation of the exciton eigenstate. At $U = U_{\rm c}$, we find $\psi_{\mathcal{A}\downarrow;\mathcal{B}\downarrow}(\bm{k})=\psi_{\mathcal{B}\uparrow;\mathcal{A}\uparrow}^{*}(\bm{k})=0$ [Fig.~\ref{wf:evolution}(d)].
 The  magnetically ordered state emerges as  a superposition of the nonmagnetic ground state of the band insulator and 
states containing multiple coherent excitons. The choice of the  phase factor carried by each condensing exciton, equal to $0$ or $\pi$, corresponds to the  Z$_2$ time-reversal symmetry breaking ($M>0$ or $<0$ on either sublattice). 
Note that the magnetic moments on the two sublattices must
be opposite because  of the relative ``$-$'' sign between  $\psi_{\sigma; \sigma} ({\bm r}_1 \in \mathcal{A},{\bm r}_2\in \mathcal{A})$ and $\psi_{\sigma; \sigma} ({\bm r}_1 \in \mathcal{B},{\bm r}_2\in \mathcal{B})$, i. e., the system develops
 N{\'e}el magnetic ordering for $U>U_c$.

The time-reversal symmetry is restored as  $U$ decreases further below $U_c$ (in the band insulator), and  the wave function of the exciton becomes extended [see  Figs.~\ref{wf:evolution}(a)--\ref{wf:evolution}(c)], as expected from the  reduction of the binding energy.
For zero binding energy  ($\zeta=0$), the integral
$\frac{1}{{\cal N}_{u}}\sum_{\bm{k}}\psi_{\mathcal{A}\downarrow;\mathcal{B}\downarrow}(\bm{k})e^{i\bm{k}\cdot(\bm{r}_{2}-\bm{r}_{1})}$
that determines the real space wave function $\psi_{\downarrow\downarrow}(\bm{r}_{1}\in \mathcal{A},\bm{r}_{2}\in \mathcal{B})$ is dominated by the singular points $\bm{k}_{0}$ that minimize the
particle-hole excitation energy $2\sqrt{\delta^{2}+b_{\bm{k}}^{2}}$.
In the simplest case where there is a unique singular point $\bm{k}_{0}$,
$\psi_{\downarrow\downarrow}(\bm{r}_{1}\in \mathcal{A},\bm{r}_{2}\in \mathcal{B})\propto\frac{1}{{\cal N}_{u}}e^{i\bm{k}_{0}\cdot(\bm{r}_{2}-\bm{r}_{1})}$
takes the form of a plane wave. 

\bibliography{ref}
\end{document}